\documentclass[aps,pre,twocolumn,groupedaddress,superscriptaddress,showpacs]{revtex4-1}

\usepackage{amsmath}
\usepackage{graphicx}
\usepackage{float}
\usepackage{color}

\begin{document}

\title{Field-driven dynamical demixing of binary mixtures} 

\author{Andr\'e S. Nunes}
\affiliation{Departamento de F\'{\i}sica, Faculdade
de Ci\^{e}ncias, Universidade de Lisboa, P-1749-016 Lisboa, Portugal, and
Centro de F\'isica Te\'orica e Computacional, Universidade de Lisboa,
P-1749-016 Lisboa, Portugal}

\author{Akshat Gupta}
\affiliation{Departamento de F\'{\i}sica, Faculdade
	de Ci\^{e}ncias, Universidade de Lisboa, P-1749-016 Lisboa, Portugal, and
	Centro de F\'isica Te\'orica e Computacional, Universidade de Lisboa,
	P-1749-016 Lisboa, Portugal}

\author{Nuno A. M. Ara\'ujo}
\email{nmaraujo@fc.ul.pt}
\affiliation{Departamento de F\'{\i}sica, Faculdade
de Ci\^{e}ncias, Universidade de Lisboa, P-1749-016 Lisboa, Portugal, and
Centro de F\'isica Te\'orica e Computacional, Universidade de Lisboa,
P-1749-016 Lisboa, Portugal}

\author{Margarida M. Telo da Gama}
\affiliation{Departamento de F\'{\i}sica, Faculdade
de Ci\^{e}ncias, Universidade de Lisboa, P-1749-016 Lisboa, Portugal, and
Centro de F\'isica Te\'orica e Computacional, Universidade de Lisboa,
P-1749-016 Lisboa, Portugal}

\begin{abstract}
We consider mixtures of two species of spherical colloidal particles that differ in their hydrodynamic radii, but are otherwise identical, in the presence of an external field. Since the particle-particle and particle-field interactions are the same for both species, they are completely mixed in the thermodynamic limit in the presence of any static field. Here, we combine Brownian Dynamics and Dynamic Density Functional theory of fluids to show that for sufficiently large differences in the hydrodynamic radius of the particles (and corresponding differences in their electrophoretic mobilities) dynamical demixing is observed.  These demixed states are transient but, under certain conditions, packing effects compromise the relaxation towards the thermodynamic states and the lifetime of the demixed phases increases significantly.
\end{abstract}

\maketitle

\section{Introduction}

The ability to grow colloidal structures from their spontaneous self-organization is among the most critical challenges of the Physics of Soft Condensed Matter \cite{Whitesides2002, Lu2013}. Successful strategies include  induced crystallization \cite{Everts2016, Wu2009, Haxton2015, Hatch2016}, the use of surfaces and interfaces, with and without  templates \cite{Blaaderen1997, Cadilhe2007, Ramsteiner2009, Darshana2016}, particle functionalization \cite{Wang2012, Bianchi2011, Araujo2017} and external fields \cite{Volpe2014, Nunes2016}. For single component systems, the feasibility of the targeted structures controlled through the response of the colloidal system to external perturbations, e.g., changes in temperature, interaction with the surface, or with the external fields \cite{Lowen2008, Dias2013}. For multi-component systems, asymmetries in the dynamical response are expected to lead to richer dynamics, paving the way to new routes for the assembly of the target structures \cite{Bechinger2016, Weber2016, Kumari2017}.

As a step in this direction, we investigate the dynamics of field-driven self-organization of binary mixtures.  Spatially dependent fields are used in experiments as virtual molds to drive colloidal suspensions into structures that exhibit the symmetries of the field \cite{Demirors2013}. The final structure of each species will depend on the particle-field and particle-particle interactions. In order to focus on the role of the dynamics, we consider binary mixtures of particles that differ solely in their hydrodynamic radius, but have identical particle-field and particle-particle interactions. Thus, in the corresponding thermodynamic phases, both species are completely mixed and the spatial arrangement of the particles corresponds to that of a single component system with the same (total) density, as the species are indistinguishable from the thermodynamic point of view. 

We use Brownian Dynamics (BD) and Density Functional Theory of fluids (DDFT), to study the dynamics of such mixtures in the presence of an external field. We consider a field corresponding to a sinusoidal potential with a minimum at the center of the simulation box. We show that when the field is switched on, albeit both species  tend to move towards the minimum, dynamical demixing is observed for sufficiently different hydrodynamic radii, since the species respond differently to the field. We also find that, as the field leads to an increase in the density at the minimum of the potential, at sufficiently high densities packing effects may compromise the access to the thermodynamic (mixed) phases.

The paper is organized in the following way. In Section \ref{sec::model}, we describe the details of the model and simulations. Results from particle-based simulations (BD) and continuum theory (DDFT) are discussed in Section \ref{sec::results}. Finally, we draw some conclusions in Section \ref{sec::conclusions}. 

\begin{figure*}[ht]
	\centering
	\includegraphics[width=14cm]{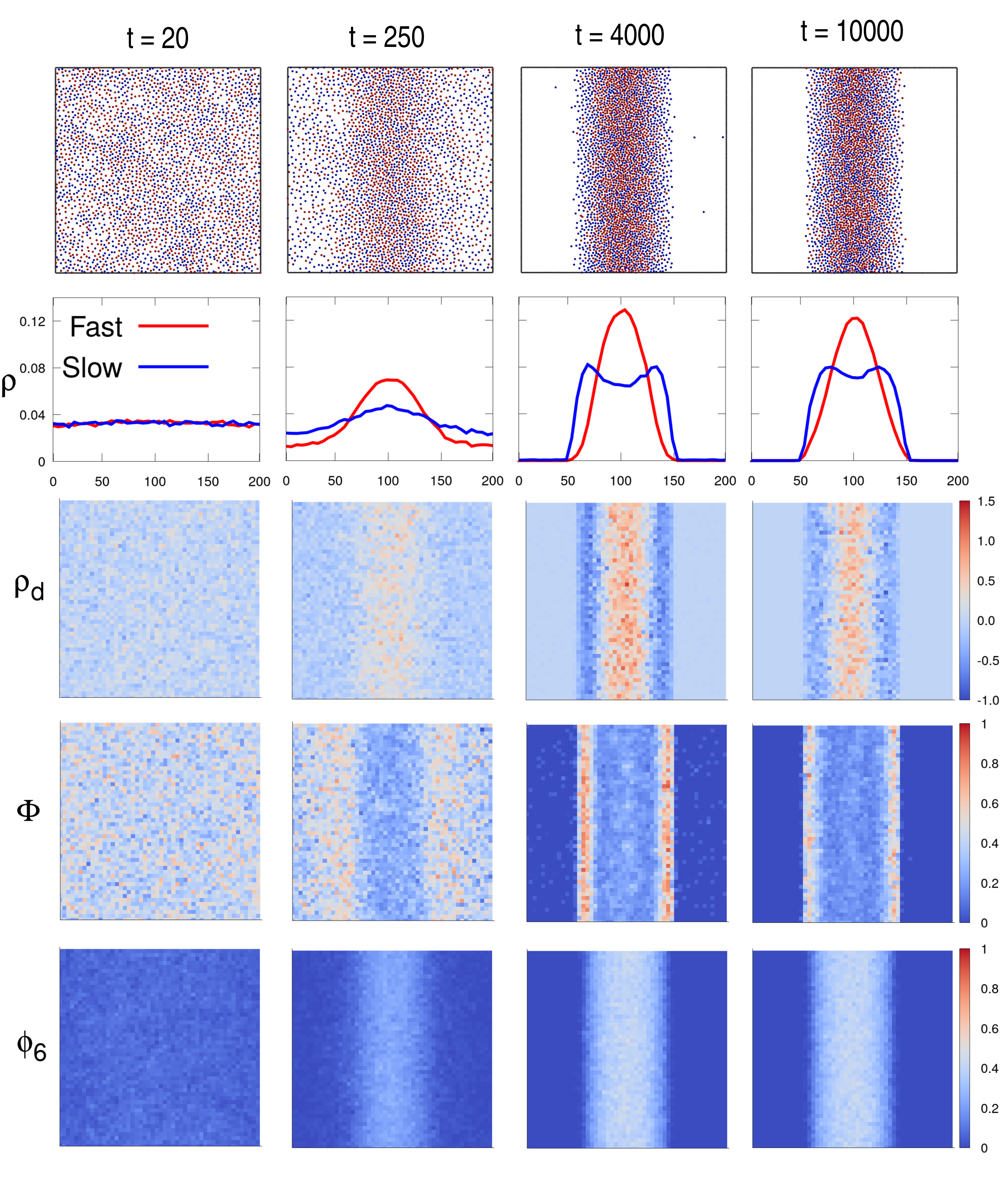} 
	
	\caption{In the first row are snapshots at $t=20,250,4000,10000$. Blue and red particles are slow and fast, respectively. The ratio between the friction coefficients is $\Gamma=6$ and the energy scales are $V_E=70$ and $V_0=50$, respectively. The second row shows the density profiles along the $x$-direction. In the third row we depict the difference in the local densities in each region of the simulation box obtained by discretizing the simulation space using a mesh of squares with lateral size $4r_p$. The fourth row depicts the local segregation parameter and the fifth row the local 6-fold bond order parameters, obtained from the same discretization. The results shown in the $2^{nd}$ to $5^{th}$ rows are averages over $20$ samples.}\label{fgr:snapshothigh}
\end{figure*}

\section{Model and Simulations~\label{sec::model}}
We consider a a binary mixture of colloidal particles in a two-dimensional square domain of lateral size $L$ and periodic boundary conditions in both directions. The equation of motion of a  particle $i$ in the overdamped regime is
\begin{equation}
\gamma_i \frac{d\vec{r}_i}{dt} = -\vec{\nabla} _i\left[\sum_{j} V_{ij}(r) + V_{ext}(\vec{r}_i)\right] + \vec{\xi}_{i},\quad j\neq i\label{Eq:MotionEq}
\end{equation}
where $\gamma_i$ is Stokes friction coefficient and $\vec{\xi}_i$ is the stochastic force that describes the fast-fluctuating particle-fluid interaction. This force is sampled from a Gaussian distribution, with zero mean and second moment $\left\langle \xi_i^k(t)\xi_i^l(t')\right\rangle = 2k_BT\gamma_i\delta_{kl}\delta(t-t')$ and thus it is uncorrelated in time and space. The indices $k$ and $l$ represent the corresponding degree of freedom, $k_B$ is the Boltzmann constant and $T$ the thermostat temperature.  We consider two species of spherical particles that differ in their hydrodynamic radii, i.e., they have different $\gamma_i$, namely, $\gamma_f$ and $\gamma_s$, with $\gamma_f<\gamma_s$. We denote the species by fast and slow, $\gamma_f$ and $\gamma_s$, respectively. From the Stokes-Einstein relation, $D_i=k_BT/\gamma_i$, fast particles have a higher diffusion coefficient than slow ones. 

We model the interparticle pairwise interaction with a Yukawa potential,
\begin{equation}
V_{pp}(r)=V_{0}\frac{\exp\left(-\alpha r\right)}{r},\label{Yukawa}
\end{equation}
where $V_0$ sets the energy scale of the interaction and $\alpha = (2r_p)^{-1}$ its range. $r_p$ is the characteristic particle radius (both for fast and slow particles) which we set as the unit of length. For simplicity, we consider only repulsive interactions ($V_0>0$). Note that the potential parameters are independent of the type of the particles and only depends on the distance $r=\vert\vec{r}_i-\vec{r}_j\vert$ between particles $i$ and $j$.

The external potential is sinusoidal in the $x$-direction with a minimum at the center of the simulation box
\begin{equation}
V_{ext}(x) = V_E \cos\left( \frac{2\pi}{L}x \right),\label{ExternalField}
\end{equation}
where $V_E$ sets the strength of the interaction between the field and the particles. Note that the potential only varies in $x$-direction. The initial configuration is random with the particles uniformly distributed. When the external field is switched on, particles of both species are driven towards the center of the simulation box (along the $x$-direction). Since the interparticle and particle-field interaction potentials are independent of the particle species, the particles in the thermodynamic phases are completely mixed. 

The potential is expressed in units of $k_BT$ and time is defined in units of the Brownian time $\tau=r^2_p\gamma (k_BT)^{-1}$. Eq. \ref{Eq:MotionEq} is integrated with a numerical scheme proposed by Bra\'nka and Heyes ~\cite{Branka1999}, which consists of a second-order stochastic Runge-Kutta scheme with a time-step of $\Delta t=10^{-4}\tau$. The simulations ran for a time up to $t=10^4$. The box linear size is $L=200$ and the mixture is equimolar with a total number of particles $N=2600$. The energy scales are $V_E=70$ and $V_0=50$, unless stated otherwise.

\section{Results~\label{sec::results}}

Fast particles have higher diffusivity and electrophoretic mobility than slow ones. We define the relevant control parameter $\Gamma=\gamma_{f}/\gamma_{s}$, which is adimensional. Particles are identical in the limit $\Gamma=1$, with the relaxation dynamics and equilibrium properties of single component systems, as discussed in the Ref.~\cite{Nunes2016}. Below, we consider the case $\Gamma\neq1$, when fast and slow particles differ in their hydrodynamic radii. 

\subsection{High temperatures~\label{subsec::hightemperatures}}

First we consider the case where the temperature is high enough to avoid crystallization. Figure \ref{fgr:snapshothigh} (first row) shows snapshots at different instants in time (increasing from left to right) at $\Gamma=6$. When the field is switched on, particles are dragged towards the center of the simulation box, forming a band along the $y$-direction, reproducing the symmetry of the potential. 

The asymmetry between the particles hydrodynamic radii leads to distinct responses to the external potential. Fast particles move faster than slow ones, triggering dynamical demixing. Note that, at all times, there is a higher concentration of fast particles along the center than of slow ones. This is also observed in the evolution of the density profiles in the $x$-direction shown in the second row of Fig. \ref{fgr:snapshothigh}, where the red and the blue lines are for fast and slow particles, respectively. The third row in the same figure shows the difference between the fast and slow particle densities in each region of the simulation box, normalized by the total density, $\rho_d = (\rho_{f} - \rho_{s})/\rho_t$, where the positive regions (in red) are where the density of fast particles is higher. We see that, starting from a uniform distribution of both species, segregation is observed, with the fast particles in the center and the slow particles accumulated at the boundaries of the band.

For a more quantitative analysis, we divide the simulation box using a square mesh and define the local segregation parameter inside a mesh-cell as
\begin{equation}
	\Phi= \frac{1}{N_c}\sum_{i}^{N_c} \frac{\left(n_{il} - n_{iu} \right)^2}{\left(n_{il} + n_{iu} \right)^2},
\end{equation}
where $N_c$ is the number of particles inside the cell, $n_{il}$ and $n_{iu}$ are the number of like or unlike particles surrounding particle $i$ within a cut-off distance $r_{cut}=3$, respectively. This parameter is one if there is a single species inside the cell and zero when the number of particles of each species is the same. The time evolution of $\Phi$ is shown in the fourth row of Fig. \ref{fgr:snapshothigh}. The spatial dependence of the particle segregation is clear at $t=4000$. The boundaries of the band consist mainly of slow particles, while in the center, although there is a higher density of fast particles, the segregation is not complete, since slow particles that are initially near that region are trapped inside the band. At later times, $\Phi$ decreases at the boundaries of the band due to the expected thermodynamic mixing.

We also measured the bond order parameter, $\phi_6$, for a particle $i$ defined as 
\begin{equation}
\phi_6 = \frac{1}{6}\left|\sum_{j}^{N_b}e^{i6\theta_{ij}}\right|,
\end{equation}
where $N_b$ is the number of neighbors surrounding a particle within the cut-off radius, $r_{cut}$, and  $\theta_{ij}$ is the angle between a line connecting particles $i$ and $j$ and the $x$-direction. The bond order parameter is one when the particles are organized in a hexagonal structure. As shown in the last row of Fig. \ref{fgr:snapshothigh}, for the considered model parameters, we find no six-fold symmetry inside the band and the particles are in a fluid-like state.

\subsection{Continuous model~\label{subsec::continuosmodel}}

Particles from both species are identical from the thermodynamic point of view. So, they should be completely mixed in the thermodynamic limit. That is why we see some degree of mixing at later times in  Fig. \ref{fgr:snapshothigh}. But the dynamics of mixing is very slow and to observe complete mixing much longer simulations are needed. To achieve much longer timescales and analyze the dynamics of mixing, we considered here a simple, coarse-grained continuum model based on Dynamic Density Functional Theory (DDFT) of fluids.

\begin{figure}[ht]
	\centering
	\includegraphics[width=\columnwidth]{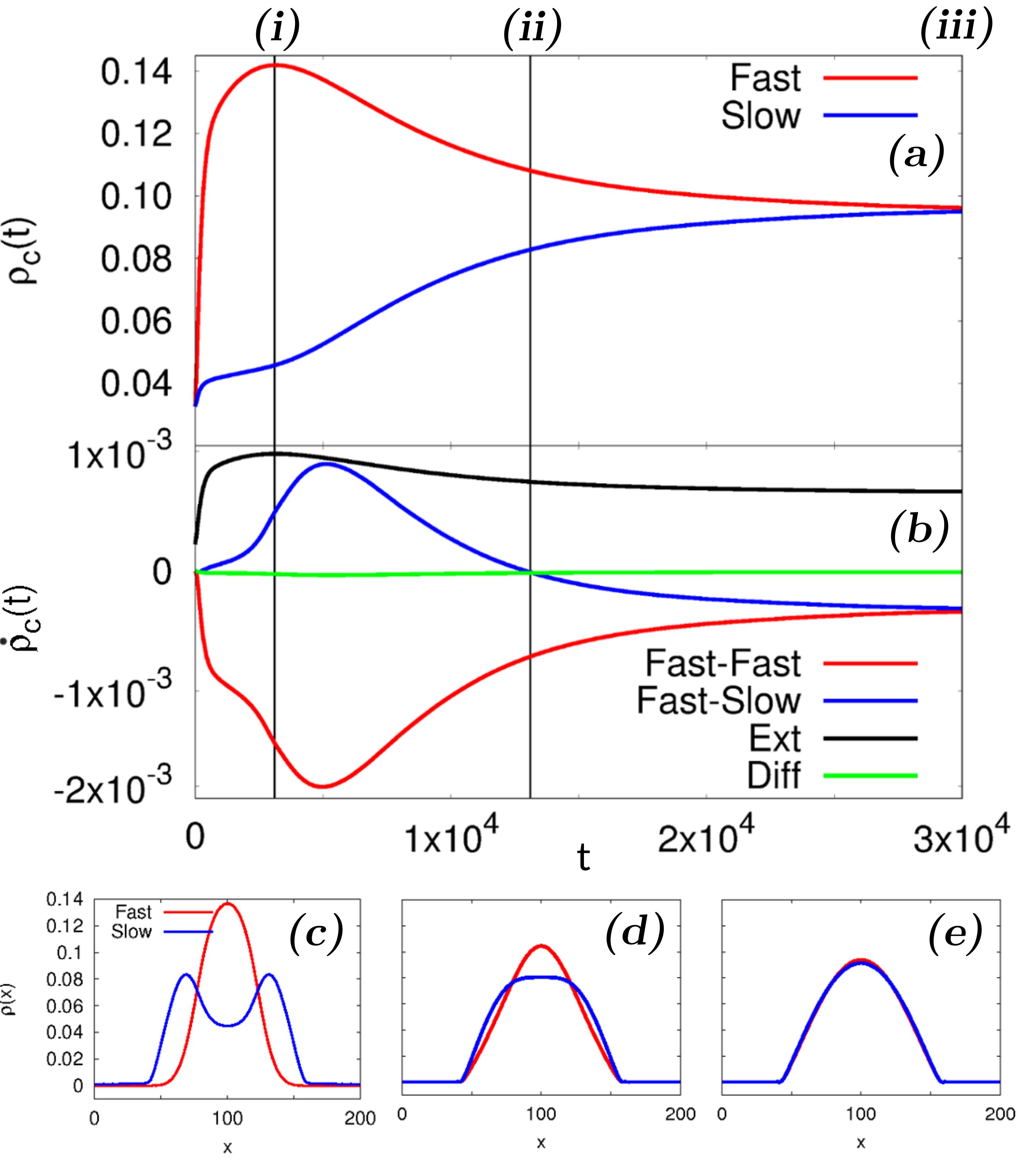}
	\caption{(a) Density evolution in the center of the band obtained from DDFT, with the red line depicting the fast particle density and the blue line the slow particle density. (b) Evolution of the individual terms of Eqs. \ref{fast}. The red line represents the interaction between the fast particles, the blue line the interaction between different particles, the black line the interaction with the external potential and the green line the diffusion. The black vertical lines separate the three different regimes. (c), (d) and (e) show the density profiles at the boundaries between the three regimes (i), (ii) and (iii), respectively.}\label{fgr:DFT_densityEvol}
\end{figure}

For the system under consideration, the free energy functional is written in terms of the fast and slow densities as
\begin{equation}
	\begin{split}
		&\mathcal{F}[\rho_f,\rho_s]=\\ 
		&\sum_{i=f,s}\int k_{B}T\rho_i(\vec{r})\left[\log\left(\rho_i(\vec{r})\Lambda^{2}\right)-1\right]d\vec{r} +\\
		&\sum_{i=f,s}\sum_{j=h,c}\frac{1}{2}\int\int \rho_i(\vec{r}) \rho_j(\vec{r}') V_{pp}(\vec{r}-\vec{r}')d\vec{r}'d\vec{r} +\\ 
		&\sum_{i=f,s}\int\rho_i(\vec{r})V_{ext}(\vec{r})d\vec{r},
	\end{split} 
	\label{funtional}
\end{equation}
where $\Lambda$ is the de Broglie wavelength and the indices $f$ and $s$ stand for fast and slow. The first term is the ideal gas contribution to the free energy, the second describes the interaction between the particles, and the third is the contribution from the interaction with the external potential. We derived the equation for the time evolution of the densities as~\cite{Tarazona1999}
\begin{equation}
\gamma_i \frac{\partial\rho_i(\vec{r},t)}{\partial
	t}=\nabla.\left[\rho_i(\vec{r},t)\vec{\nabla}
\frac{\delta\mathcal{F}[\rho_f(\vec{r},t),\rho_s(\vec{r},t)]}{\delta\rho_i(\vec{r},t)}\right],
\label{DDFT}
\end{equation}
obtaining two non-linear diffusion equations,
\begin{equation}
	\gamma_f\frac{\partial\rho_f}{\partial t}=\nabla.\left[A\rho_f\vec{\nabla}\rho_f+A\rho_f\vec{\nabla}\rho_s+\vec{\nabla}V_\mathrm{ext}\rho_f\right]+k_BT\nabla^2\rho_f
	\label{fast}
\end{equation}
and
\begin{equation}
	\gamma_s\frac{\partial\rho_s}{\partial t}=\nabla.\left[A\rho_s\vec{\nabla}\rho_s+A\rho_s\vec{\nabla}\rho_f+\vec{\nabla}V_\mathrm{ext}\rho_s\right]+k_BT\nabla^2\rho_s,
	\label{slow}
\end{equation}
where $A=\int V_{pp}(\vec{r})d\vec{r}$ is the interaction parameter in the local density approximation, see Ref.~\cite{Nunes2016} for further details. In this approximation the densities are considered smooth functions of the position, $\rho(\vec{r}')\approx\rho(\vec{r})$, where $\vec{r}'$ is a point in the vicinity of $\vec{r}$.

We solved Eqs. \ref{fast} and \ref{slow} using the finite elements method \cite{comsol} and the evolution of the densities in the center of the band is shown in Fig. \ref{fgr:DFT_densityEvol} (top). Both densities increase until the point where the density of fast particles reaches a maximum and then starts to decrease. In the asymptotic limit both densities tend to the same value that corresponds to a completely mixed state, as expected in the thermodynamic limit. We can also analyze how each term in Eqs. \ref{fast} and \ref{slow} contributes to the dynamics. The middle plot in Fig. \ref{fgr:DFT_densityEvol} shows the detailed evolution of Eq. \ref{fast} with the contribution from the fast-fast particle interaction (first term in the square brackets, red line in the plot), from the fast-slow particle interaction (second term in the square brackets, blue line in the plot), from the fast particles with the external potential (third term in the square brackets, black line in the plot) and finally from the diffusion term (last term of the equation, green line in the plot). These results show that there are three dynamical regimes, delimited by the black vertical lines in the plot. In the first regime, there is dynamical demixing caused by the difference in mobility of the two species that culminates with the peak density of the fast particles. At the maximum, the density profile of the slow particles is bimodal, as shown in Fig. \ref{fgr:DFT_densityEvol}(c). In the second regime, there is mixing where the fast particles density profile decreases due to the diffusion and the interaction with particles of the same species. Also, the slow particles density profile in the center increases at a fast rate. At the end of this regime the density profile of the slow particles is no longer bimodal, shown in Fig. \ref{fgr:DFT_densityEvol}(d). In the third regime, the density profiles slowly converge to the completely mixed state.

\subsection{Low temperatures~\label{subsec::lowtemperatures}}
\begin{figure}
	\centering
	\includegraphics[width=3.5cm]{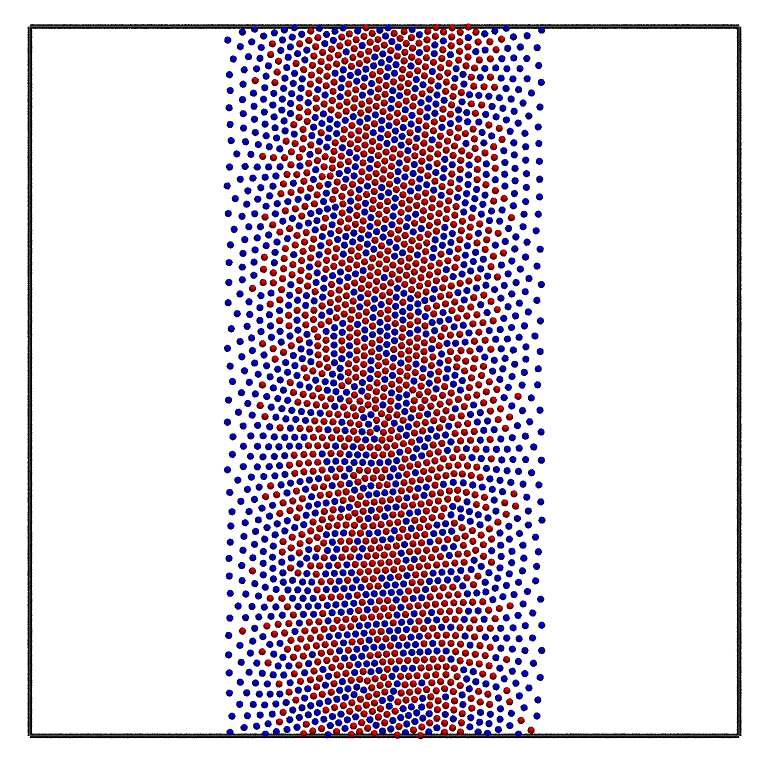} 
	\includegraphics[width=4.2cm]{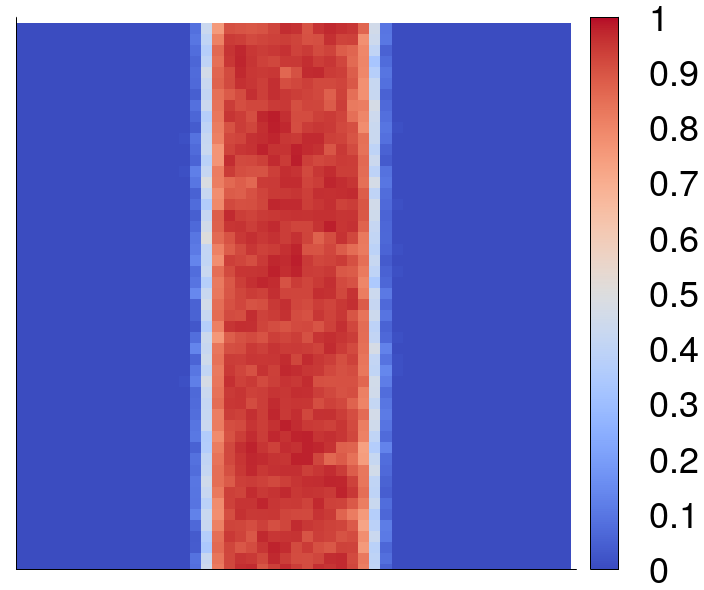}
	\caption{Asymptotic state obtained at $t=10000$ (left) at low temperatures showing crystallization. Red particles are fast and blue are slow. The ratio between diffusion coefficients is $\Gamma=6$ and the external and particle-particle potential scales are $V_E=7000$ and $V_0=5000$, respectively. On the right-hand side we show the local 6-fold bond order parameter at the end of the simulation, averaged over $20$ samples. }\label{fgr:snapshotlow}
\end{figure}

As discussed above, the demixing is dynamic and therefore transient. For long enough times, the system is expected to reach an equilibrium configuration,  where both particle species have the same density profile. However, if the mixing time of the third regime is of the order of the experimental time scale, in practice the dynamical demixed states are the observed ones. The symmetry of the external potential favors the accumulation of particles along the $y$-direction, in the center of the simulation box. At high enough densities (and low temperatures), due to the local increase of the density, the particles rearrange in a crystalline-like structure. Mixing becomes much slower in such regimes, due to packing effects.

Figure \ref{fgr:snapshotlow} shows the final state when the temperature is decreased by two orders of magnitude. From the system snapshot (left) at late times as well as from the bond order parameter (right) it is clear that, the entire band crystallizes and we can, therefore, effectively freeze the system in the demixed state.

\subsection{Dynamical demixing~\label{subsec::dynamical demixing}}

\begin{figure}
	\centering
	\includegraphics[width=\columnwidth]{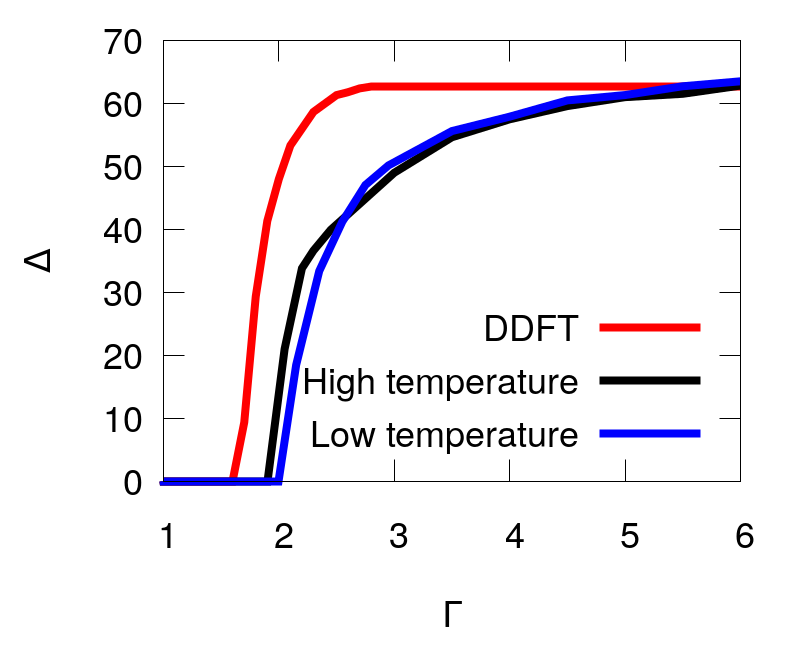}
	\caption{The distance, $\Delta$, between the two peaks in the density profile of slow particles at high temperatures (black), low temperatures (blue) and DDFT (red). Segregation only occurs for high $\Gamma$'s.}\label{fgr:delta}
\end{figure}

The demixing occurs because the two particle species have different mobilities. Since the dynamical demixing is caused by the difference in hydrodynamic radii, we now study its dependence on $\Gamma$. We define a parameter, $\Delta$, that measures the distance between both peaks in the slow density profile at the time when the density of fast particles in the center is maximal (at the end of the first regime). The results from Fig. \ref{fgr:delta} show that there is a threshold $\Gamma^*$ below which no demixing is observed, corresponding to $\Gamma^*\sim1.6$ from DDFT and $\Gamma^*\sim2$ in the simulations. At high  enough $\Gamma$ the value of $\Delta$ saturates since the number of particles in the simulation box is finite, imposing an upper limit to the width of the band. 

Note that there are little differences between the curves at high and low temperatures, which is to be expected since $\Delta$ is measured at the end of the first regime before the mixing starts at high temperatures and the density profiles are similar to those in the frozen state. The differences between the thresholds in the DDFT and the simulations is due to the local density approximation, which neglects short and long distance correlations between the particles. Also, if we were to increase the number of particles in the system, keeping the density constant, the simulation density profiles (and $\Delta$) would be less sensitive to the number of particles and, in this limit, we would expect to find a sharper $\Delta$ curve, like that obtained from DDFT. 

\section{Conclusions~\label{sec::conclusions}}
We studied binary systems where the particle species are distinguished only by different electrophoretic mobilities. In the presence of space dependent external fields, the species are driven to the zero-field regions faster or slower according to their mobilities, driving a temporary demixing as the system evolves towards the thermodynamic equilibrium state. The final configuration is characterized by a completely mixed state since the energy landscape is the same for both species.  

The amount of demixing can be controlled by tuning the ratio between the friction coefficients of the two species. This state is a transient and mixing will eventually occurs. However, it is possible to increase the mixing time-scale by lowering the temperature and induce crystallization in the high density regions, in which case, the demixed state is effectively frozen.

Previous approaches to self-assembled mixtures of colloidal particles considered structures in thermodynamic equilibrium that replicate the symmetries of the virtual molds produced by external fields. However, we have shown that the particle dynamical properties play a critical role in this self-assembly, as the evolution of the system is easily trapped in arrested states for long periods of time. This represents an obstacle that needs to be suppressed if the goal is to reach the equilibrium state.  

Our results also show that the dynamics can be explored as a new method to control the self-assembly in any system where the particles have different mobilities. Adjusting the time-scales for the relaxation and arresting, one can take advantage of the different particle dynamical properties to assemble structures that, albeit not in thermodynamic equilibrium, are still reproducible and robust over a long period of time. 

\section{Acknowledgements}
We acknowledge financial support from the Portuguese Foundation for Science and Technology
(FCT) under Contracts nos. EXCL/FIS-NAN/0083/2012, UID/FIS/00618/2013, IF/00255/2013 and SFRH/BD/119240/2016.
\bibliography{refs}

\begin{thebibliography}{24}
\bibitem [{com()}]{comsol}%
  \BibitemOpen
  \href@noop {} {\enquote {\bibinfo {title} {Comsol multiphysics software is
  used to implement the fem.}}\ }\BibitemShut {NoStop}%
\end{thebibliography}
\end{document}